\documentclass[twocolumn,prb,aps,amsmath,superscriptaddress,showpacs]{revtex4}
\usepackage{float}
\usepackage{epsfig}

\def\gee        {\epsilon}
\def\go         {\omega}
\def\la         {\langle}
\def\ra         {\rangle}

\renewcommand{\[}{\left[}
\renewcommand{\]}{\right]}
\renewcommand{\(}{\left(}
\renewcommand{\)}{\right)}

\restylefloat{figure}

\begin{document}
\title{First--principles calculation of the plasmon 
resonance and of the reflectance spectrum of Silver in the $GW$ approximation}
\author{Andrea Marini}
\affiliation{   
Istituto Nazionale per la Fisica della Materia,
Dipartimento di Fisica dell'Universit\`a di Roma ``Tor Vergata''}
\author{Rodolfo Del Sole}
\affiliation{
Istituto Nazionale per la Fisica della Materia,
Dipartimento di Fisica dell'Universit\`a di Roma ``Tor Vergata''}
\author{Giovanni Onida}
\affiliation{
Istituto Nazionale per la Fisica della Materia,
Dipartimento di Fisica dell'Universit\`a di Roma ``Tor Vergata''}
\affiliation{
Istituto Nazionale per la Fisica della Materia,
Dipartimento di Fisica dell'Universit\'a di Milano,
Via Celoria 16, I--20133 Milano, Italy}
\date{\today}

\begin{abstract}
 We show that the position and width of the plasmon resonance in
 Silver are correctly predicted  by ab--initio calculations including
 self--energy effects within the  $GW$ approximation.
 Unlike in simple metals and semiconductors, quasiparticle corrections
 play a key role and are essential to obtain Electron Energy  
 Loss in quantitative agreement with the experimental data.
 The sharp reflectance  minimum at $3.92$\,eV, that cannot be 
 reproduced within DFT--LDA, is also well described within $GW$. 
 The present results solve two unsettled 
 drawbacks of linear response calculations for Silver.
\end{abstract}

\maketitle


\section{Introduction}
\label{sec:intro}

Although the coexistence of quasiparticles and collective 
excitations in an interacting system is well known, our knowledge of 
their properties and mutual interactions in real materials
is far from being complete.
One of the most successful approaches to the calculation of the
quasiparticle (QP) band structure for a wide range of materials is the GW 
method~\cite{hedin65}.
Starting from a non--interacting representation of the system, electrons 
and holes are screened by the surrounding electronic clouds created 
through the Coulomb interaction.
This description has been shown to be rather general in describing 
direct and  inverse  photoemission spectra, also for complicated systems 
as reconstructed surfaces and clusters~\cite{gwrev,surfclu}.
Noble metals, on the other hand, have been studied only very recently.
Full quasiparticle calculations have been carried out only for  
Cu obtaining an excellent agreement with the experimental 
band structure~\cite{CUPRL}.
Collective modes, i.e. plasmons, occur at energies for which the real part of the
dielectric function vanishes with a corresponding small imaginary part; 
they can be observed experimentally as
sharp peaks in Electron Energy Loss Spectra (EELS).  A well established
technique to calculate EEL spectra uses the single particle or random phase approximation
(RPA) for the polarization function, obtained in terms of Ab--Initio
energy bands.  The latter can be calculated within the Kohn
Sham (KS)~\cite{KS}  formulation of Density Functional Theory~\cite{DFT}
in the Local Density Approximation~\cite{LDA} (DFT--LDA) or in the
Generalized Gradient Approximation (GGA)~\cite{GGA}, when the
exchange--correlation potential $V_{xc}$ is considered as an approximation to the
self--energy operator $\Sigma$. This approach completely neglects
self--energy corrections, that is the difference
$\Sigma-V_{xc}$, and excitonic effects, whose inclusion has been
faced only recently. In the case of silicon, Olevano and
Reining~\cite{si} showed that using the quasiparticle energy bands without
including excitonic effects the shape of the plasmon peak worsens
with respect to the experiment.
Inclusion of self--energy corrections and excitonic effects yields a spectrum very
similar to DFT--LDA one and to the experiment. Ku and
Eguiluz~\cite{KU} obtained a correct positive dispersion of the plasmon
width in K using the single particle approximation, with  no many--body corrections
beyond DFT--LDA. 
In these cases many-body effects
(quasiparticle corrections and/or excitonic effects) are not required to
describe correctly the experimental data. These results agree with
the general feeling that excitonic effects partially cancel self--energy
corrections.  A similar result has  been found for Cu~\cite{AngelCU,CUPRB}, where
the RPA response function calculated without many-body
corrections yields good agreement with the experimental EEL and optical spectra.

In this framework the case of Silver is rather surprising: the
experimental EELS is dominated by a sharp plasmon peak at $3.83$\,eV~\cite{herenreich},
whose position and width are badly reproduced in DFT--LDA RPA~\cite{cazalilla}. In
particular, a width of about $0.5$\,eV is obtained within this approach,
to be contrasted with a much narrower experimental width ($\sim\,100$\,meV).
A similar discrepancy occurs in the reflectance spectrum, where a
very narrow dip at $3.92$\,eV is hardly reproduced by DFT--LDA calculations.
Some papers have recently appeared~\cite{Lee,Zhukov} which correct
DFT--LDA results by empirical scissors-operators shifts (or similar),
meant to better account for the band structure. Improved dielectric
functions are obtained in this way, but no solution of the puzzles
mentioned above has been given.  

In Sec.~\ref{sec2} we show that non trivial
quasiparticle corrections on the highly localized {\it d} bands
strongly modify the absorption spectrum threshold. As a consequence,
in Sec.~\ref{sec3} we show that
the plasmon position and width are renormalized by quasiparticle effects 
leading them close to the experimental values. 
Also the reflectance dip at $3.92$\,eV turns out to be well described in the
same scheme.
Finally in Sec.~\ref{sec4} we summarize our conclusions.

\section{Quasiparticle effects on the absorption spectrum}
\label{sec2}
We proceed calculating the band structure of Silver within DFT--LDA. The
diagonalization of the KS hamiltonian is performed using norm-conserving
pseudopotentials (PPs) and a plane waves basis~\cite{CUPRB}. The use of
soft (Martins-Troullier~\cite{MT}) PPs allows us to work at full
convergence with a reasonable kinetic energy cutoff (50 Ry.). Many-body
corrections are added on top of the DFT--LDA band structure
following the implementation of the GW method
described in Ref.~\cite{CUPRL}. The resulting quasiparticle band
structure of Ag at high symmetry points is compared with DFT--LDA
results in Table~\ref{tab1}. While the deeper energy levels remain mostly
unchanged, a downward shift of about $1.3$\,eV of the top {\it d} bands
leads to a decrease of the bandwidth, and hence to an excellent agreement
with experiment. 
The residual discrepancies between the position of the GW bands and the 
experimental results ($\sim$\,0.2\,eV) is larger than in the case of copper~\cite{CUPRL}
($\sim$\,0.05\,eV). This is a consequence of the QP renormalization of the Fermi level
in silver ($\sim$\,0.6\,eV), not present in copper, which yield 
the large QP corrections reported in Table~\ref{tab1} and an additional error.
Differently from semiconductors, GW
corrections {\it do not} act as a rigid shift of the whole occupied band
structure with respect to the empty (conduction) part. 
QP corrections are highly non trivial since even their
sign turns out to be band/{\bf k}--point dependent, as we have already
shown in the case of copper~\cite{CUPRL}.
\begin{table}[H]
\begin{tabular}{lcccc} \hline\hline
 &   & DFT--LDA  & $GW$ &  Experiment  \\ \hline
Positions 	& $\Gamma_{12}$ 		& $-3.57$ & $-4.81$ &  $-4.95$ \\
of        	& $X_5$         		& $-2.49$ & $-3.72$ &  $-3.97$ \\
{\it d}-bands   & $L_3(2)$         		& $-2.71$ & $-3.94$ & $-4.15$ \\ \hline
        	& $\Gamma_{12}-\Gamma_{25'}$ 	& $1.09$ & $0.94$ & $1.11$ \\
Widths  	& $X_5-X_3$ 			& $3.74$ & $3.39$ & $3.35$ \\
of      	& $X_5-X_1$ 			& $3.89$ & $3.51$ & $3.40$ \\
{\it d}-bands 	& $L_3(2)-L_3(1)$		& $1.98$ & $1.85$ & $1.99$ \\
        	& $L_3-L_1$ 			& $3.64$ & $3.17$ & $2.94$ \\ 
        	& $X_5-X_2$ & $0.27$ & $0.29$ & $0.38$ \\  \hline\hline
\end{tabular}
\caption{\footnotesize{
Theoretical band widths and band energies for silver, at
high-symmetry points. GW energies are relative to the QP Fermi
Level. The striking agreement with the experimental results shows that the
silver band--structure is very well described
at the $GW$ level. The values in the last column are taken from
Ref.{\protect\cite{fuster}}  where spin--orbit splittings have been removed 
by making degeneracy--weighted averages. }} \label{tab1}
\end{table}

\begin{figure}[H]
\begin{center}
\epsfig{figure=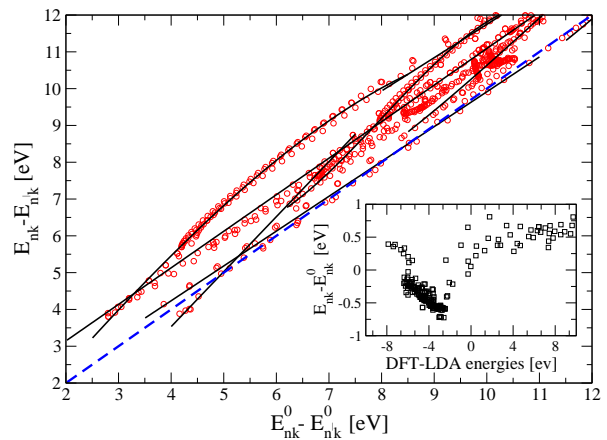,clip=,bbllx=60,bblly=25,bburx=590,bbury=725,angle=-90,width=8cm}
\end{center}
\caption{\footnotesize{GW optical--transition energies (circles) plotted
as a function of DFT--LDA ones for a regular grid of 29 {\bf k}--points in the
irreducible Brillouin zone. The dashed line corresponds to vanishing GW
corrections while full lines are our fitted curves. They are used to
extend the GW corrections to the optical transition energies of a larger
{\bf k}--points grid. GW corrections shift upward the electron--hole energies,
except for some transitions between $4$ and $5$\,eV (as
discussed in the text). In the inset GW corrections are shown versus
DFT--LDA energies.}}
\label{fig0}
\end{figure}

To calculate the EEL spectra, we use the QP band structure 
to evaluate the inverse dielectric function $\gee^{-1}\(\go\)$~\cite{lfe}:
\begin{align}
 \gee^{-1}\(\go\)=\[\gee_{ib}\(\go\)-\frac{\go_D^2}{\go\(\go+i\eta\)}\]^{-1},
\label{eq1}
\end{align}
where 
$\gee_{ib}\(\go\)$ is the interband contribution and 
$\go_D=9.48$\,eV is the Drude plasma frequency, both calculated 
Ab--Initio following the procedure described in Refs.~\cite{CUPRB,drude}.
The interband RPA dielectric function is 
given by 
\begin{multline}
 \gee_{ib}\(\go\)=
 1-4\pi\lim_{{\bf q}\rightarrow {\bf 0}}
 \int_{BZ} \frac{d^3 {\bf k}}{\(2 \pi\)^3} \\
 \sum_{n\neq n'} \frac{\left|\la n'{\bf k-q}|e^{-i{\bf q}\cdot{\bf r}}|n {\bf k}\ra\right|^2}
      {\left|{\bf q}\right|^2}
 \frac{f_{n',{\bf k-q}}-f_{n,{\bf k}}}{\go+E_{n,{\bf k}}-E_{n',{\bf k-q}}+i\eta},
\label{eq2}
\end{multline}
where $0 \leq f_n\({\bf k}\)\leq 2$ represents the occupation number 
summed over spin components. $\eta=0.1$\,eV has been used in the present calculations. 
The ${\bf q}\rightarrow {\bf 0}$ limit of Eq.\,(\ref{eq2}) has been done including
the effects of the pseudopotential non--locality, as done
in Ref.~\cite{CUPRB}.
The GW optical-transition energies 
$\(E_{n,{\bf k}}-E_{n',{\bf k}}\)$ are obtained fitting the QP 
corrections calculated at 29 {\bf k}--points of a regular grid in the irreducible 
Brillouin zone (BZ). 
Separate fitting curves (shown in Fig.~\ref{fig0}) 
have been used for selected band pairs $\(n,n'\)$ and {\bf k} regions in the BZ, in
order to correctly reproduce the energy dependence of the quasiparticle
corrections. 
From Fig.~\ref{fig0}  it is 
evident that GW corrections are highly non--trivial (ranging from $-0.75$
to $0.5$\,eV for filled states, and from $0$ to $0.75$\,eV for empty states, as
shown in the inset). GW optical-transition energies are generally above
the dashed-line, representing the condition of vanishing quasiparticle corrections
$\(E_{n,{\bf k}}-E_{n',{\bf k}}\)=\(E^0_{n,{\bf k}}-E^0_{n',{\bf k}}\)$,
and lead to upward shifts of the electron--hole energies.
However between $4$ and $5$\,eV some electron--hole energy shifts are
negative. These correspond to transitions close to the L point from the
Fermi level to the 7th (empty) band.

\begin{figure}[H]
\begin{center}
\epsfig{figure=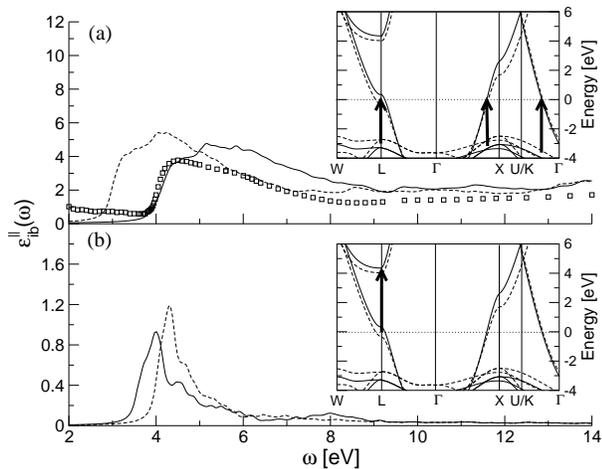,clip=,bbllx=40,bblly=6,bburx=605,bbury=735,angle=-90,width=8cm}
\end{center}
\caption{\footnotesize{Interband absorption spectrum $\gee^{''}_{ib}\(\go\)$ of Ag
calculated using the quasiparticle GW band structure. Theoretical
spectra do not contain the intraband contribution. In both panels and
insets: solid line, GW; dashed line, DFT--LDA.  
Only in panel\,(a):
boxes, full experimental $\gee^{''}\(\go\)$ {\protect\cite{palik}}. 
In panel\,(a) all the transitions
except those involving the {\bf k}--points near the L point are included.
These are considered separately in 
panel\,(b).
In the insets the DFT--LDA band structure is compared with the $GW$ result,
and the arrows indicate the most important transitions involved
near the absorption threshold.
In panel\,(b) quasiparticle corrections shift some transition
energies at the L point {\it below} the main threshold of panel\,(a).
These are responsible for the plasmon damping.}}
\label{fig1}
\end{figure}

In Fig.~\ref{fig1} the GW interband contribution $\gee^{''}_{ib}\(\go\)$
is compared with the DFT--LDA result. In the upper panel all the transitions
except those involving the {\bf k}--points near the L point are included: GW
corrections shift the whole spectra to {\it higher} energies, with the
threshold energy occurring at $\sim 4$\,eV, in agreement with experiment.
In panel\,(b) we consider the transitions (sketched in the inset) not included in panel\,(a).
At difference with
panel (a), GW shifts the DFT--LDA spectrum toward {\it lower} energies.
In conclusion, weak interband transitions start at the threshold in panel\,(b) ($3.36$\,eV),
while the main onset of $\gee^{''}_{ib}\(\go\)$ is that of panel\,(a), $\sim 4$\,eV.
The resulting (total) dielectric function is shown in the inset of
Fig.~\ref{fig2}.

\section{The renormalized plasmon frequency}
\label{sec3}
\begin{figure}[H]
\begin{center}
\epsfig{figure=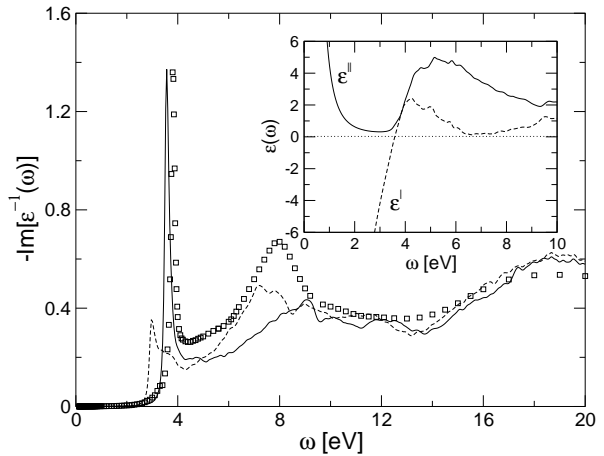,clip=,bbllx=70,bblly=40,bburx=575,bbury=695,angle=-90,width=8cm} \end{center}
\caption{\footnotesize{
Electron Energy Loss Spectrum (EELS) of Silver. Solid line: GW.
Dashed line:
DFT--LDA. Boxes: experiment {\protect\cite{palik}}. The non trivial
quasiparticle GW corrections improve considerably the DFT--LDA plasmon
peak, yielding a striking agreement with the experiment. The $GW$
dielectric function is shown in the inset. }}
\label{fig2}
\end{figure}

The two different effects of GW corrections on $\gee^{''}_{ib}\(\go\)$
shown in Fig.~\ref{fig1} are crucial in determining the properties of the
plasmon resonance whose frequency, $\go_p$, is defined through
the relation 
\begin{align}
 \gee_{ib}\(\go_p\)-\frac{\go_D^2}{\go_p^2}=0.
\label{eq4} 
\end{align}
$\go_p$ is, in general, complex since decay into electron--hole
pairs gives to the plasmon a width 
proportional to $\gee^{''}_{ib}\(\go_p\)$. Now, the sharp onset of
$\gee^{''}_{ib}\(\go\)$ at $\sim 4$\,eV (see inset of Fig.~\ref{fig2}) is
responsible for a solution of Eq.\,(\ref{eq4}) at $3.56$\,eV,  just below the
main interband threshold.
Within DFT--LDA, instead, the interband threshold of $\gee^{''}_{ib}\(\go\)$ at
$\sim 3$\,eV remains below $\go_p$, resulting in a strongly damped
plasmon peak as shown in Fig.~\ref{fig2} (dashed line). From
Fig.~\ref{fig1}, panel\,(b), we see that the only transitions contributing
to the plasmon width, proportional to $\gee^{''}_{ib}\(\go_p\)$, are
those from the Fermi surface to the first conduction band at the L
point. These decay channels are in agreement with many experimental
results~\cite{wooten} that, using temperature and alloying techniques,
have singled out the electron--hole transitions responsible for the plasmon
damping.

GW-corrected EELS is compared with DFT--LDA and experimental results in
Fig.~\ref{fig2}: the plasmon peak, underestimated in intensity and
position in DFT--LDA, is shifted toward higher energies and strongly
enhanced by GW corrections, in striking agreement with experiment.
In Fig.~\ref{fig1} an overestimation of the absorption spectrum with 
respect to the experiment is observed (similar to that found in Cu~\cite{CUPRB}).
This is reflected in an underestimation of the loss intensity close to $8$\,eV, and
might be due to the neglection of QP renormalization 
factors in Eq.\,(\ref{eq2}) and to excitonic dynamical effects~\cite{rod97}. 
Despite this partial agreement in the $8$\,eV region, the present {\it Ab--initio} GW
calculation is able to reproduce correctly the plasmon resonance.
This resonance can be interpreted as a
collective (Drude like) motion of electrons in the partially filled
band. However, according to Eq.~\ref{eq4}, $\omega_p$ does not coincide
with the bare Drude frequency $\omega_D$, the difference arising from
the screening of the electron-electron interaction by virtual interband
transitions.  The plasmon resonance, although blueshifted with respect to
DFT--LDA, remains {\it below} the main interband threshold, but overlaps
the weak low--energy tail of interband transitions 
(see Fig.~\ref{fig1}, Frame\,(b)) acquiring a small --yet finite-- width.
The highly non--trivial QP shifts of interband transitions are crucial
to obtain this result.

The polarization of the ``medium'' where
the plasmon oscillates (the {\it d} electrons) is hence important to
determine its energy and width. This polarization is absent in the homogeneous electron gas
because there are no localized {\it d} orbitals and no
interband transitions; it is weak in semiconductors (like Si), 
because interband transitions occur at energies far from that
of the plasma resonance.
The same polarization effect is present, but destructive in copper due
to the lower onset of interband transitions.
EELS peaks occur {\it above} this onset and are therefore strongly broadened.
In conclusion, the delicate interplay of plasmon--frequency renormalization with
the shift of the interband--transition onset, both due
to QP corrections, may yield (in Silver) or may not yield (in Copper) a sharp
plasmon resonance.

\begin{figure}[H] 
\begin{center}
\epsfig{figure=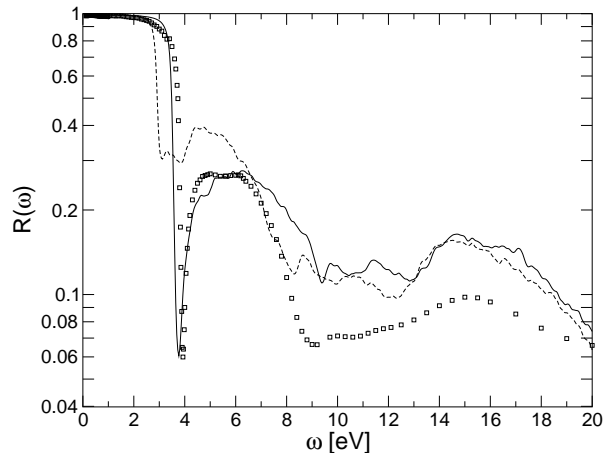,clip=,bbllx=70,bblly=40,bburx=570,bbury=695,angle=-90,width=8cm} 
\end{center} 
\caption{\footnotesize{Reflectivity
spectrum of Silver. Solid line: GW. Dashed line: DFT-- LDA. Boxes:
experiment {\protect\cite{palik}}. The experimental sharp dip at $3.92$\,eV is correctly
reproduced by $GW$, with a substantial improvement on the DFT--LDA spectrum. }}
\label{fig3} 
\end{figure}

Another important quantity is the reflectance, 
$R\(\go\) =\(\left|N\(\go\)-1\right|/\left|N\(\go\)+1\right|\)^2$, 
where N is the complex refraction index defined by  $\[N\(\go\)\]^2= \gee\(\go\)$.
In Fig.~\ref{fig3} we compare the  $GW$  $R\(\go\)$ with the DFT--LDA one, 
and with  experimental results~\cite{palik}.  The latter shows a
very narrow dip at $3.92$\,eV, close to the plasmon frequency, arising from the
zero-reflectance point $\omega_0$, defined as $\epsilon\(\omega_0\)=1$.
Again, the width and depth of this reflectance dip are related to the
imaginary part of $\epsilon(\omega)$.
GW corrections make $\omega_0$ to occur below the main onset of
interband transitions, and hence produce a very narrow and deep
reflectance minimum. Here the agreement between $GW$ results and
experiments for the intensity and width of the dip at $3.92$\,eV is even
more striking than in the EELS.

This result is of great importance for the optical and EELS properties
of Ag {\it surfaces}. Very recent calculations of Reflectance
Anisotropy Spectra (RAS) within DFT-- LDA for Ag(110)~\cite{pat} were not able to
reproduce quantitatively a sharp dip observed experimentally. This feature,
at an energy  $\go_r\ = 3.8$\,eV, has been assigned to a bulk resonance,
arising in the RAS spectrum when  $\gee\(\go_r\)=1$.  Hence, it
is the same occurring in the reflectance spectrum of Fig.~\ref{fig3}.
Its width and shape are strongly related to the reflectance dip of Fig.~\ref{fig3},
and therefore they need GW corrections to be well reproduced.

\section{Conclusions}
\label{sec4}
We have performed a calculation of the EEL and
reflectivity spectra of silver within the RPA approach, using the
quasiparticle band structure calculated within the 
Ab--Initio GW method.  We have shown that the peculiar, well known
plasmon peak in the EELS and the deep reflectivity minimum observed
experimentally, are quantitatively described by theory, for the first
time without requiring adjustable parameters.  Our theoretical calculations
do not contradict previous results for semiconductors and simple metals
(where many-body effects are less important) if the role played by {\it
d} orbitals is correctly interpreted. 

This work has been supported by
the INFM  PRA project ``1MESS'', MURST-COFIN 99 and
by the EU through the NANOPHASE Research Training
Network (Contract No. HPRN-CT-2000-00167).
We thank Angel Rubio and Lucia Reining for useful discussions.


\begin{thebibliography}{100}
\bibitem{drude}		The Drude plasma frequency $\go_D$ has been 
			calculated using the DFT--LDA states. This is justified
			as the DFT wavefunctions are not renormalized
			by the GW corrections and the dispersion of the	metallic  
			band close to the Fermi level is poorly affected by QP
			effects.
\bibitem{lfe}		Local field effects are not included in the present 
			work. However, their influence for 
                        noble~\cite{AngelCU,CUPRB} 
			and transition metals~\cite{gwrev} has been extensively
			studied and found negligible on both EELS and absorption 
			spectra.
\end{thebibliography}
\end{document}